\documentclass{evn2004}
\usepackage{txfonts}
\usepackage{graphicx}
\begin{document}
   \title{From truck to optical fibre: the coming-of-age of eVLBI}

   \author{Arpad Szomoru\inst{1}
          \and
          Andy Biggs\inst{1}
          \and
          Mike Garrett\inst{1}
          \and
          Huib Jan van Langevelde\inst{1}
          \and
          Friso Olnon\inst{1}
          \and
          Zsolt Paragi\inst{1}
          \and          Steve Parsley\inst{1}
          \and
          Sergei Pogrebenko\inst{1}
          \and
          Cormac Reynolds\inst{1}
          }

   \institute{Joint Institute for VLBI in Europe, Postbus 2, 7990 AA Dwingeloo, The Netherlands
             }
             
             \abstract{ Spurred by the advent of disk-based recording
               systems and the nearly explosive increase of internet
               bandwidth, eVLBI (Parsley et al. \cite{parsley}) has
               undergone a remarkable development over the past two
               years. From ftp-based transfers of small amounts of
               astronomical data, through near real-time correlation
               (disk-buffered at the correlator), it has culminated
               this spring in the first three telescope real-time
               correlation at JIVE (Onsala, Westerbork and Jodrell
               Bank).  In this paper we will give a review of this
               development and the current state of affairs. We will
               also address the current limitations and the way we may
               improve both bandwidth and reliability and finally we
               will discuss the opportunities a true high-bandwidth
               real-time VLBI correlator will provide.  }

   \maketitle
%
\section{Introduction}

At this time, the replacement of Mk{\sc IV} magnetic tape recorders with
Mark5 units, PC-based disk recorder systems (Whitney \cite{whitney}), is in full swing.  As a
direct result of this development, eVLBI, connecting stations through
Mark5 units via optical fibres, has become possible and may well
become the successor to disk-based recording.

Fibre communication networks are ideally suited to the real-time
transfer of huge amounts of data over long distances. The adoption of
direct fibre connections by eMERLIN signals the progress that is
being made in this area. As large networks are being deployed for the
use of research in Europe, and as networks become more flexible, the
introduction of a real-time VLBI system is a realistic goal to pursue.

The EVN and JIVE are actively involved in a proof-of-concept (POC) programme which
aims to demonstrate the feasibility of real-time VLBI using IP routed
networks. This POC is supported by G\'{E}ANT, a collaboration between
European National Research and Education Networks, the European
Commision, and Dante (Delivery of Advanced Network Technology to
Europe). The aim is to connect at least four EVN telescopes
to the correlator at JIVE, ultimately at datarates of 1~Gbps, and feed
these data streams directly into the correlator.

In connecting the telescopes, local loops (the last mile connection)
are a critical item. At the time of writing, the Westerbork, Onsala
and Torun radio telescopes have 1~Gbps connections, while Jodrell Bank
and Arecibo are connected through 155~Mbps. The upgrade of Jodrell
Bank to 2.5~Gbps and the connection of Medicina at 1~Gbps are planned
for the end of this year.

JIVE itself was connected to Netherlight in September 2002, and
currently can make use of 6 lambda's, each capable of carrying 1~Gbps,
and one direct Gbps connection to the Westerbork Synthesis Radio
Telescope.

The outline of this paper is as follows. In Section~2 we discuss
transport protocols and possible optimizations. Section~3 deals with
tests and results, and in Section~5 we consider possible implications
for the operation of the EVN. Section~5 lists the conclusions.


\section{Tcp versus udp, tuning issues}

In principle, data can be transferred in many ways and forms,
especially on dedicated networks. Here we exclusively consider IP
based protocols using existing networks. 

The most widely used internet transfer protocols are TCP and UDP. Both
are based on the IP protocol, that is, both make use of IP packets,
consisting of data and a header.  However, while the TCP protocol
tries to maximize reliability using a direct connection, through
three-way handshakes, acknowledgments, backoff algorithms and
re-sending of packets in the case of packet loss, UDP is
connectionless and will simply send off packets without further
accounting. Apart from the case of dedicated point-to-point
connections, this makes UDP faster than TCP, but large amounts of data
may go missing, and packets may even arrive in the wrong order.
Although a certain amount of data loss is acceptable in the case of
VLBI, UDP can only be used in combination with a mechanism that would
count and re-order packets, replacing gaps with dummy data.
 
Different protocols, like Scalable and High Speed TCP, have been and
are being developed, attempting to combine high speed with
high reliability.  For the time being, the software provided with the
Mark5 units only supports TCP and UDP, although a VSI-E software
module, currently under development at Haystack, will provide more
flexibility (D. Lapsley, private communication).

The performance of TCP can be improved, sometimes quite dramatically,
by the tuning of several parameters. In current Linux implementations
many of these parameters are set to default values that are inappropriate
for high-speed networks.

\begin{figure*}
\includegraphics[angle=0, width=0.5\textwidth]{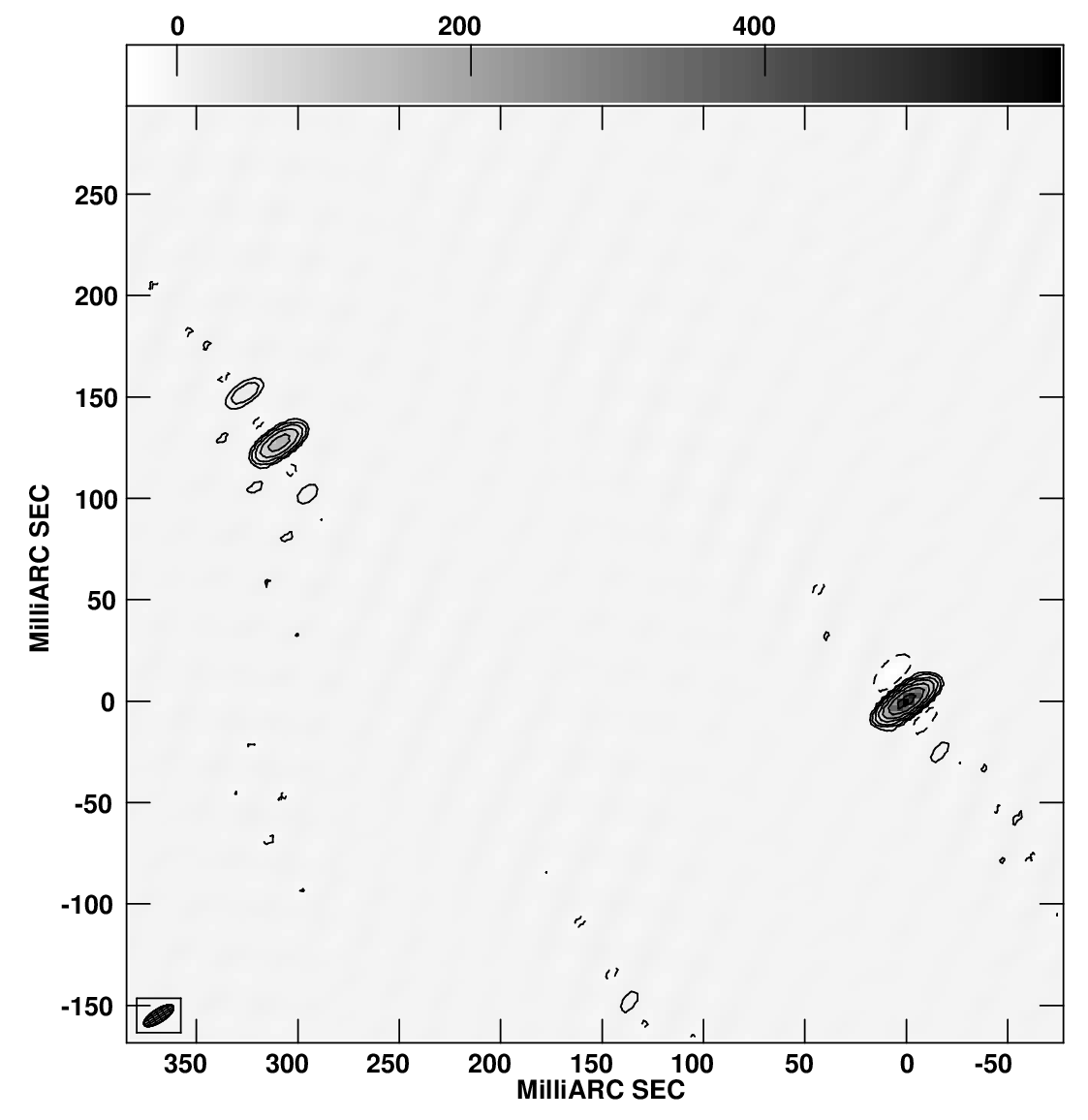}
\includegraphics[angle=0, width=0.5\textwidth]{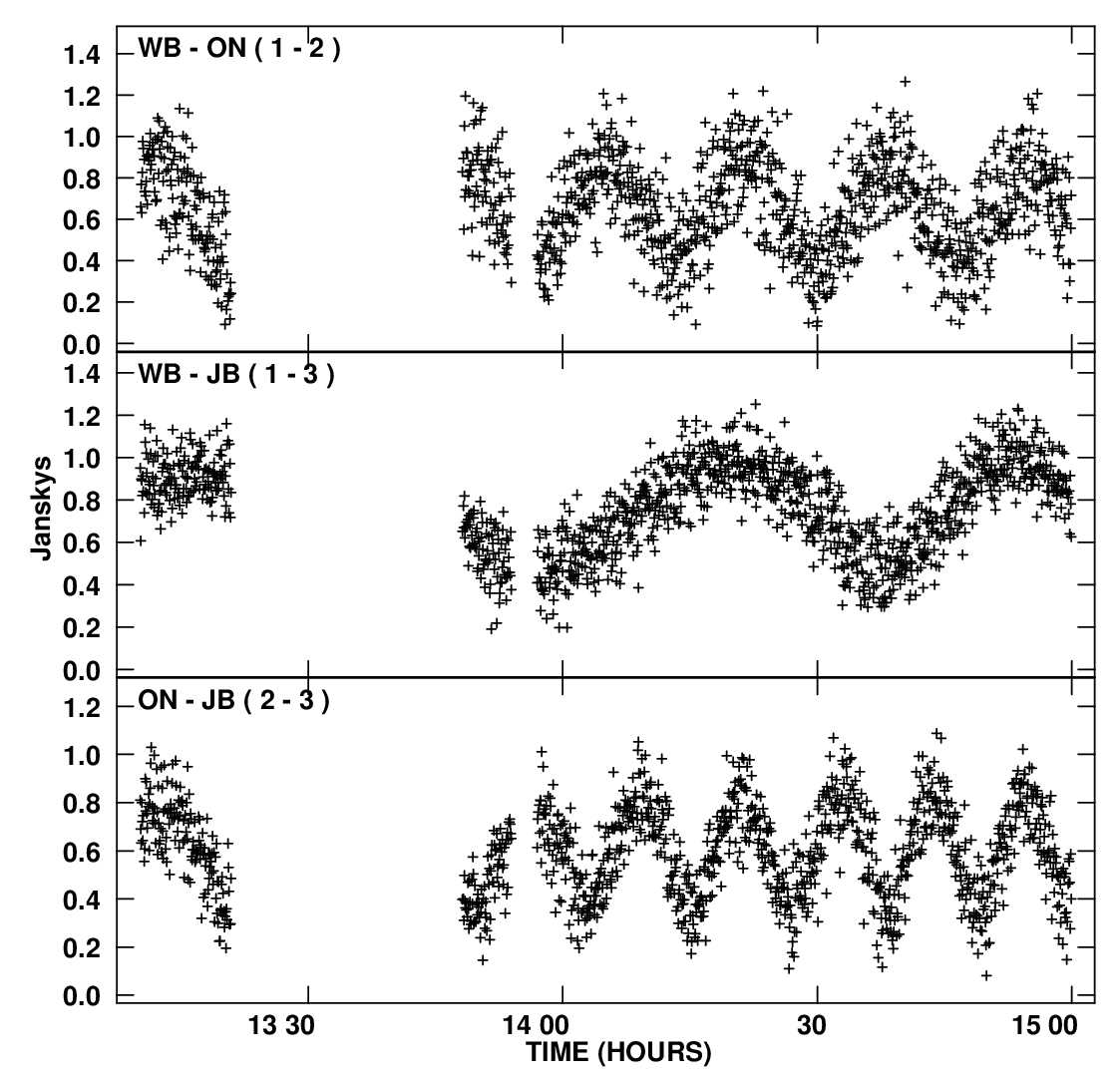}
\caption{First real-time eVLBI map of the
gravitational lens system, JVAS B0218+357 (left), two images of
a bright radio quasar, separated by 334mas across the sky, result
in a beating correlated amplitude on all 3 baselines, the period
depending on the baseline length and orientation. 
}
\label{evlbimap}
\end{figure*}

\begin{itemize}
  
\item TCP window or congestion window. This buffer determines how many
  packets can be sent (and received) at one time and is calculated
  as the product of the roundtrip time and the bandwidth. To enable
  large window sizes, the default system buffer sizes must be
  adjusted.
  
\item Txqueuelen--netdev\_backlog. These determine the sizes of queues
  between kernel network subsystems and the netword interface card
  driver, on sending and receiving ends. Here too, default sizes must
  be increased.

\item Selective acknowledgment (sack). The implementation of sack in
  current Linux releases can cause a serious delay in connection recovery
  after a burst of packet loss, and should be disabled.
  
\item MTU size. The use of jumbo frames (9000 bytes instead of 1500)
  has a dramatic effect on throughput, but all equipment (such as
  switches and routers) along the path must support this. Generally
  this is not (yet) the case.
  
\item Interrupt moderation/coalescence. With every TCP packet generating one
  interrupt, some motherboard/cpu combinations are overwhelmed by the
  number of interrupts during Gbps transfers. It is possible to limit the
  number of interrupts generated by adjusting the parameters of the
  network interface card driver, thereby reducing the load on the cpu. This
  does however increase the delay.

\end{itemize}

\section{Tests and results}

In this section we review the tests done over the past few years at
JIVE and the results they have led to.  These tests were conducted
both on the bench, between Mark5 units connected through patch cables
or via the switch at Netherlight, and between Mark5 units in Dwingeloo
and at various observatories. In some cases different PCs (not
Mark5s) were also used to measure throughput across international
networks. We have done both memory-to-memory tests (using iperf and
similar programs) and tests involving real data being read from and to
recording media. Network stress tests were also done, in which as much
data as possible was sent simultaneously from several stations to
JIVE.

All of the tuning parameters mentioned in the previous section influence
transfer speeds to some degree, and some of them influence each other.
Some of the parameters are only important at high speeds, which can
only be reached in memory-to-memory tests on the bench.  Transfer of
real data can be influenced by yet other factors, like the load on the
PCI bus of the PC or the speed of disk access. The best result we have
obtained so far was the transport of real data via a patch cable at a
rate of $\sim$500~Mbps, with jumbo frames, and $\sim$300~Mbps without.  Note that
memory-to-memory transfers over such a connection will easily reach
speeds of $\sim$900~Mbps, even without jumbo frames (if properly tuned).

The results of these tests are summarized in Table~1. In the
following I will describe various transfer modes.

\begin{table*}
  \caption[]{Test results of TCP and UDP transfers in various setups. Listed are the maximum
data rates in Mbps. Note that in many cases transfer rates were not constant over time, and moreover 
were not symmetrical, i.e. data rates from and to JIVE would differ.
}
  \label{tcpudptest}
  $$
  \begin{array}{p{0.25\linewidth}lllllll}
    \hline
    \multicolumn{1}{c|}{\ } & \multicolumn{2}{c|}{\rm Mem-Mem} & \multicolumn{2}{c|}{\rm Disk2net-Net2disk}  
& \multicolumn{1}{c|}{\rm In2net-Net2disk} & \multicolumn{1}{c}{\rm In2net-Net2out} \\
\hline
     \multicolumn{1}{c|}{\ } & {\rm udp} & \multicolumn{1}{l|}{\rm tcp} & {\rm udp} & \multicolumn{1}{l|}{\rm tcp} & \multicolumn{1}{l|}{\rm tcp} & {\rm tcp} \\
  \hline
  \noalign{\smallskip}
  Bench via patch      &     & 930 & 250 &     &     & 256 \\
  idem, jumbo frames   &     & 960 &     & 544 &     & 512 \\
  Bench via Amsterdam  & 500 & 360 &     &     &     & 256 \\
  idem, jumbo frames   &     &     & 341 & 456 &     &     \\
  Westerbork-Jive      & 867 & 680 &     &     & 256 &  64 \\
  idem, jumbo frames   &     &     & 249 & 378 &     &     \\
  Bologna-Jive         & 670 & 128 &     & 307 &     &     \\
  Jodrell-Jive         &  50 &  70 &     &     &  64 &  32 \\
  Arecibo-Jive         &     &  88 &     &     &     &  32 \\
  Torun-Jive           & 800 & 260 &     &     &     &  32 \\
  Onsala-Jive          &     &     &     & 177 & 256 &  64 \\
  Jive-Haystack        & 612 &     &     &  71 &     &     \\
  \noalign{\smallskip}
  \hline
\end{array}
$$
\end{table*}

\subsection{ftp-based eVLBI}

One of the first practical implementations of eVLBI was the transfer
of astronomical data through ftp for the purpose of fringe tests. For
this, small amounts of data are transferred from Mark5 diskpacks to
normal Linux files and ftp'd to JIVE. At first these data were then
again stored on Mark5 diskpacks and correlated in the regular way,
nowadays the software correlator package developed and maintained by
the Radio Astronomy Group in Kashima Space Research Center is used,
leaving the hardware correlator available for production correlation.
This method has greatly improved the response time of JIVE to report
technical problems to the observatories and made the EVN a more
reliable instrument.

As this method is based on standard ftp, the bandwidth of the
connection is not a critical factor. However, the amount of data that
can be transferred in this way remains limited (many stations still
have poor connectivity), and the correlation itself is time consuming.
Typically, a transfer (some 15 to 30 seconds worth of data, up to
~1GB/station) will take of the order of minutes to hours, while the
correlation itself (single baselines to one station, normally
Effelsberg) takes about one hour (on an 8-node dual 2GHz Opteron Linux
cluster).

\subsection{Dual buffered eVLBI}

The next step was to bypass the stage of storing data as
Linux file by using the Mark5 commands disk2net and net2disk. Although
not very different from regular ftp, this removes the need for extra
storage and reduces the time involved. Again, bandwidth is not
critical and in fact this way of operation could cut down on operating
costs by eliminating the need for transporting diskpacks. The
disadvantage is that during the transfer, which generally will be
slower than real-time, the Mark5 units are not available for either
recording or playback.

Several tests were done in this mode, testing connectivity
and data integrity, using telescopes at Westerbork and Onsala and the
Medicina Mark5 unit that had been hooked up to a Gbps connection at
the GARR POP in Bologna.

\subsection{Single buffered eVLBI}

The last step leading up to real-time eVLBI was to stream data
directly from the formatters at the stations (Mark5 command in2net) to
diskpacks at JIVE. On January $15^{\rm th}$ 2004 this mode was tested
with Westerbork, Onsala and Cambridge (via Jodrell). Due to the slower
link to Jodrell, data from Cambridge were sent only at 64~Mbps, and
higher rates were recorded and transferred overnight.  Rates of up to
256~Mbps were tested. Good fringes were obtained at 256~Mbps between
Onsala and Cambridge, and on all baselines at 128~Mbps. The fringes
were produced within half an hour of the end of the observation, the
final image within 24 hours.

\subsection{Real-time eVLBI}

Once it became possible to servo net-streaming data in the Mark5
units, we could go ahead and implement real-time eVLBI. As telescopes
are not available at all times for testing, a number of tests were
done using a simulation setup.  This setup consisted of three Mark5
units, one unit (acting as formatter) playing back pre-recorded data
into a second unit, which sends the data across a patch cable to a
third unit, which plays the data into the correlator. In this way we
first of all could establish feasibility, and in the second place we
could determine the highest possible transfer rate under optimal
circumstances. The highest rates achieved in this way were 512~Mbps
with jumbo frames, 256~Mbps without.

In normal operations the correlator clock is set to the observing time
which is recorded in the observing logs. This observing time is
checked against the time codes in the data stream.  Software
modifications to the correlator control code were needed to
synchronize the correlator clock to UT (minus a few seconds).  The
logistics involved in connecting various telescopes were not always
trivial and resulted in a large number of long-distance telephone
calls.  Nevertheless, at this time several successful tests have been
conducted (Figure~1), with ever-increasing ease of operation.

The most recent experiment took place September 10$^{\rm th}$ of this
year, with telescopes in Arecibo, Westerbork, Torun and Cambridge
participating, and produced fringes on all baselines (Figure~2).  The
first science-driven eVLBI project is scheduled for the end of
September this year.

\begin{figure}
\includegraphics[angle=0, width=0.45\textwidth]{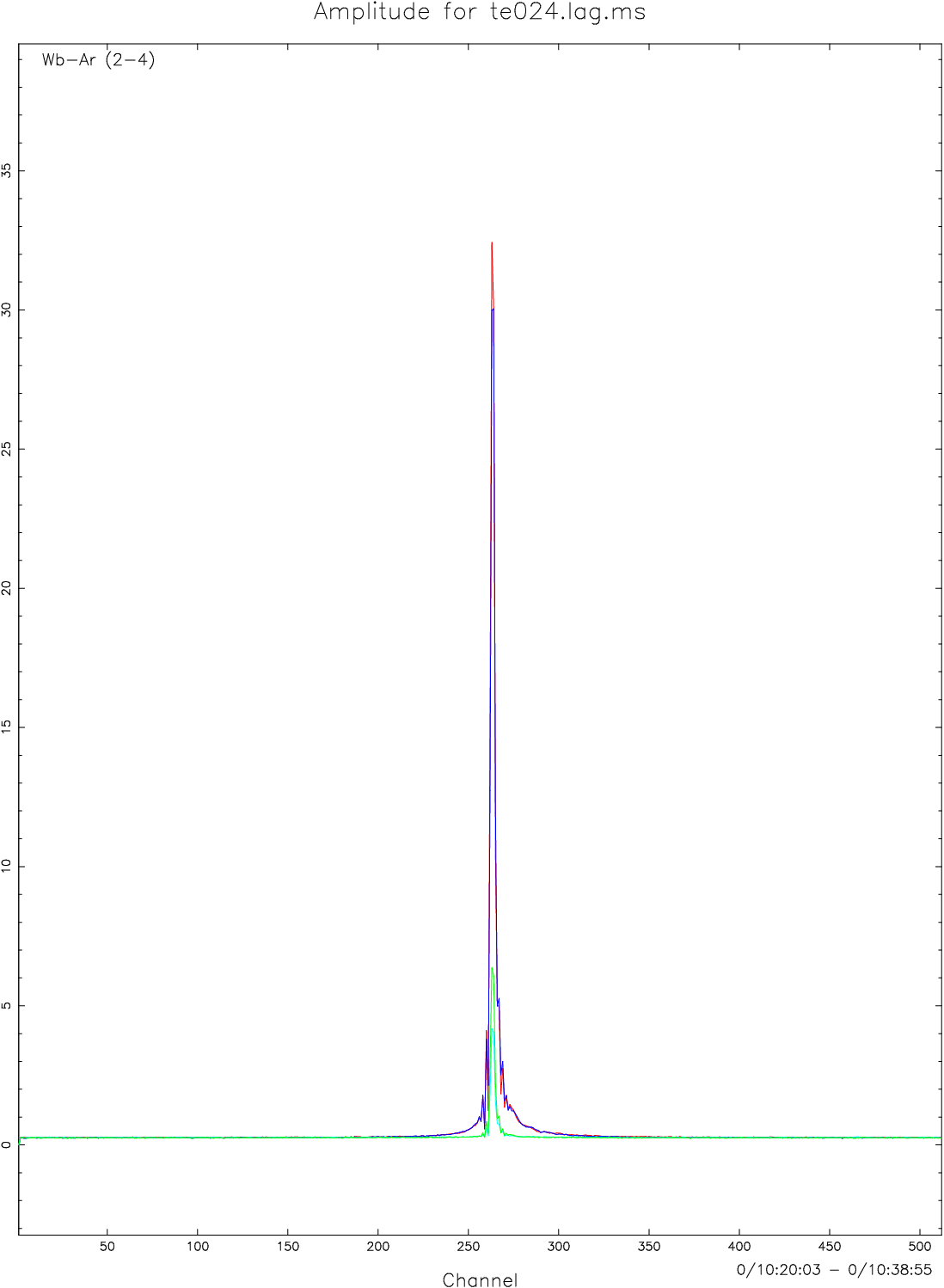}
\caption{Example of a transatlantic fringe on the Westerbork-Arecibo baseline
of 0528+134, observed September $10^{\rm th}$ 2004 using real-time eVLBI.
}
\label{evlbifringe}
\end{figure}

The highest rate we have reached so far in this mode is 64~Mbps (Onsala
and Westerbork) but because of the lower connectivity to Jodrell and
Arecibo only 32~Mbps experiments were carried out with these stations.

\section{eVLBI: towards eEVN?}

The move towards eVLBI ties in closely with other projects, all having
a tremendous impact on VLBI.  The PCInt project currently being
developed at JIVE will allow the full capacity of the correlator to be
harnessed, permitting a spectral resolution of 8092 channels per
baseline or integration times as short as 15 milli-seconds (van
Langevelde et al., these proceedings), hugely expanding the
field-of-view of VLBI. While incremental improvements in collecting
area for the EVN are expected from the Yebes-OAN 40-m telescope and
the IRA 64-m Sardinian Radio Telescope, a significant increase of
sensitivity will result from the availability of sustained data rates
of 1~Gbps and, expected on relatively short time-scales, data rates
well in excess of this.  Also in the area of improved receiver
technology, both cm and mm-VLBI are likely to see significant
progress (Garrett \cite{garrett}).

What will eVLBI mean for the EVN? Although we still are in a development phase it is
clear that the technique works and that with additional effort the
data rates can and will be cranked up. And considering the current
pace of development, both of internet bandwidth and optical
technology, one can imagine a lamda switched EVN, a network in which
telescopes are connected to the correlator through point-to-point
dedicated connections. 
\\
\\
A few advantages:

\begin{itemize}
\item No consumables. Removing the need for recording media, and its
  transport, will constitute a considerable saving of money and effort
  for the EVN. This will of course have to balance the price-tag of a
  (semi) dedicated network.
\item Fast turn-around. A real-time connected-element EVN will deliver
  data products to the users in a matter of days. This will reduce the
  long delay between conception of a project and actual research, and
  make the EVN a more exciting instrument to use. 
\item Reliability. Network performance will be monitored continuously,
  feedback in the case of problems will be nearly immediate. To
  optimise this, a flexible, more centralised form of control may be
  needed. With guaranteed reliability, it will be easier for the EVN
  to react to targets of opportunity such as GRB after-glow.
\item Future bandwidth needs. The use of standard off-the-shelf
  hardware components ensures that eVLBI will be able to take full
  advantage of commercially driven technological improvements. It
  should be noted that bandwidths beyond 1~Gbps will require a new
  correlator, which may well take a distributed form.
\end{itemize}

eVLBI will also change the current operating model of the EVN. In the
new model, the correlator will be an integral part of the network, with
monitoring information being exchanged nearly continuously between
correlator and telescopes.  As a result the EVN will become one
instrument, with highly increased coherence and reliability.  Moving
to eVLBI will enable us to combine eEVN with eMERLIN, thereby creating
the most sensitive VLBI network in the world. Further in the future,
the combination of eEVN with radio telescopes such as ALMA and SKA is
a real and exciting possibility.

\section{Conclusions}

At this point, real-time eVLBI is about to come of age and to
become, albeit on a limited scale, an EVN mode of operation in its
own right. With additional effort and investments higher data rates
with more telescopes are well within reach. In the near future,
technological developments are bound to result in an explosive growth
of available bandwidth, making 1~Gbps real-time correlation a realistic
goal.  Availability of bandwidth will no doubt create demands for yet
more bandwidth, and we should look forward to multi-Gbps correlation.
The move towards eEVN is a logical next step, and will put the EVN on
the forefront of future developments.

\begin{acknowledgements}
The European VLBI Network is a joint facility of European, Chinese, 
South African and other radio astronomy institutes funded by their 
national research councils.
\end{acknowledgements}

\end{document}